\documentclass[aps,prx,nofootinbib,twocolumn,floatfix,superscriptaddress,showpacs,longbibliography]{revtex4-1}

\usepackage{amsfonts}
\usepackage{amsmath}
\usepackage{amssymb}
\usepackage{mathtools}
\usepackage{graphicx}
\usepackage{color}
\usepackage{bbm}
\usepackage[colorlinks=true,linkcolor=blue,citecolor=blue,urlcolor=blue]{hyperref}
\usepackage{calrsfs}
\DeclareMathAlphabet{\pazocal}{OMS}{zplm}{m}{n}
\usepackage{subfigure}
\usepackage{mathptmx}
\usepackage{times,txfonts}
\usepackage{siunitx}
\usepackage{csquotes}
\usepackage{enumerate}
\usepackage{braket}

\newcommand{\ud}{\,\mathrm{d}}

\DeclareMathOperator{\tr}{tr}

\DeclarePairedDelimiterX{\infdivx}[2]{(}{)}{%
	#1\;\delimsize\|\;#2%
}
\newcommand{\infdiv}{S\infdivx}

\let\a\alpha

\let\pt\partial
\let\s\sigma

\newcommand{\op}[1]{\pmb{#1}}

\begin{document}

\title{Wigner entropy production and heat transport in linear quantum lattices}

\date{\today}

\author{William T. B. Malouf}
\email{william.malouf@usp.br}
\affiliation{Instituto de F\'isica da Universidade de S\~ao Paulo, 05314-970 S\~ao Paulo, Brazil}
\affiliation{School of Mathematical Sciences and CQNE, The University of Nottingham, University Park, Nottingham NG7 2RD, United Kingdom}

\author{Jader P. Santos}
\email{jader@if.usp.br}
\affiliation{Instituto de F\'isica da Universidade de S\~ao Paulo, 05314-970 S\~ao Paulo, Brazil}

\author{Luis A. Correa} 
\email{luis.correa@nottingham.ac.uk}
\affiliation{School of Mathematical Sciences and CQNE, The University of Nottingham, University Park, Nottingham NG7 2RD, United Kingdom}

\author{Mauro Paternostro}
\email{m.paternostro@qub.ac.uk}
\affiliation{Centre for Theoretical Atomic, Molecular and Optical Physics,
School of Mathematics and Physics, Queen's University Belfast, Belfast BT7 1NN, United Kingdom}

\author{Gabriel T. Landi}
\email{gtlandi@if.usp.br}
\affiliation{Instituto de F\'isica da Universidade de S\~ao Paulo, 05314-970 S\~ao Paulo, Brazil}

\begin{abstract}
When a quantum system is coupled to several heat baths at different temperatures, it eventually reaches a non-equilibrium steady state featuring stationary internal heat currents. These currents imply that entropy is continually being produced in the system at a constant rate. In this paper we apply phase-space techniques to the calculation of the Wigner entropy production on general linear networks of harmonic nodes. Working in the ubiquitous limit of weak internal coupling and weak dissipation, we obtain simple closed-form expressions for the entropic contribution of each individual quasi-probability current. 
Our analysis highlights the essential role played by the internal unitary interactions (node-node couplings) in maintaining a non-equilibrium steady-state and hence a finite entropy production rate.
We also apply this formalism to the paradigmatic problem of energy transfer through a  chain of oscillators  subject to self-consistent internal baths that can be used to tune the transport from ballistic to diffusive.
We find that the entropy production scales with different power law behaviors in the ballistic and diffusive regimes, hence allowing us to quantify  what is the ``entropic cost of diffusivity.''
\end{abstract}

\maketitle

\section{Introduction}

\textit{Entropy} plays a fundamental role in both thermodynamics and information theory. Unlike energy, entropy does not satisfy a continuity equation---in addition to the exchange of entropy with the environment, it can also be produced within the system. 
This additional contribution is known as entropy production and serves to gauge the irreversibility of a physical process \cite{Callen,Tome_book_15}. Indeed, one can write the following balance equation for the entropy $ S(t) $ of a system
\begin{equation}
\dfrac{\ud S(t)}{\ud t} = \Pi(t) - \Phi(t),
\label{dSdt_Pi_PHI}
\end{equation} 
where $ \Pi(t) $ is the entropy-production rate  and $ \Phi(t) $ stands for the entropy flux from the system into the environment. As a consequence of the second law of thermodynamics, $ \Pi (t) $ is always non-negative and vanishes if and only if the system is in equilibrium. When the system is allowed to relax in contact with a single heat bath, it will generally reach thermal equilibrium, so that $ \ud S/\ud t = \Pi = \Phi = 0 $. However, when it is connected to \textit{multiple baths} at different temperatures, it will instead reach a non-equilibrium steady-state (NESS) where $ \ud S/\ud t = 0 $ but $ \Pi = \Phi \geq 0 $. 
The NESS is therefore characterized by a finite entropy production rate $ \Pi $, which is continuously converted into an entropy flux $ \Phi $ and dumped into the environment.

The theory of entropy production is formulated differently depending on the type of stochastic process at hand. For classical systems, approaches based on Onsager's theory of chemical kinetics \cite{Onsager_class_Onsager_1931,Onsager_class_Machlup_1953,Onsager_class_Tisza_1957}, classical master equations 
\cite{Schnakenberg_class_1976,Schnakenberg_class_Tania_2012}, or Fokker--Planck equations \cite{Fok_Planck_class_Tome_2010,Fok_Planck_class_Spinney_2012,Fok_Planck_class_Landi_2013,Broeck_2010,Classical_th_review_Seifert_2012} have been widely used. 
Conversely, for quantum systems the problem is usually formulated in terms of Gorini--Kossakowski--Lindblad--Sudarshan (GKLS) quantum master equations,  \cite{Gorini_1976,lindblad1976generators,Spohn_1978,Leggio_Breuer_2013,Breuer_book}, repeated interactions \cite{PhysRevLett.88.097905,gennaro2008entanglement_v2,PhysRevA.76.062307,PhysRevA.87.052113,ziman2005description,uzdin2014multilevel,Barra2015,barra2017smallest,Esposito_2017,de2018reconciliation}, quantum trajectories \cite{Alexia_2017}, and fluctuation theorems \cite{Jarzynski_PRL_97,Crooks_JStatPhys_98}, among others.   
    
More recently, a formulation combining quantum phase-space methods and Fokker--Planck equations has been put forward  \cite{Santos_Wigner_2017,Brunelli_2016,Santos_spin_2018,Santos_BathView_2018} and applied to general non-thermal environments, such as squeezed, dephasing and even photon loss reservoirs, that are often used in the description of optical cavities and input-output theory \cite{Quatum_noise_gardiner}. 
This is in fact essential if one wishes to have a complete assessment  of entropy production in controlled quantum experiments. 
In addition, this framework allows to identify irreversible quasi-probability currents in phase space, which are the elementary contributions ultimately responsible for the emergence of irreversibility at the quantum level. However, the efforts so far have been focused exclusively on systems connected to a single reservoir and have not addressed the phenomenology of NESSs. In this paper, we set out to fill this gap. 

In particular, we study generic lattice systems of linearly coupled harmonic nodes connected to various heat baths. To describe the ensuing dissipative dynamics, we adopt a local master equation \cite{Wichterich_07,levy2014local}---which is accurate to lowest order in the inter-node interaction strength \cite{trushechkin2016perturbative}---and exploit its simple structure. 
We obtain closed-form analytical expressions for the steady-state irreversible entropy production and entropy flux, broken down into the elementary contributions corresponding to individual dissipation channels. 
In addition, we also obtain an expression which neatly illustrates the essential role played by the unitary dynamics in sustaining the irreversible entropy production, by enabling energy transport through the lattice. 
 
As an application, we study diffusive heat transport through a harmonic chain connected to two reservoirs at each end \cite{Znidaric_2010,Asadian,Asadian2013_QHO,Santos_Landi_16,Current_drag_09}.  
In addition, in order to switch between ballistic and diffusive behavior, we introduce auxiliary self-consistent reservoirs \cite{Bolsterli}. 
That is,  additional baths that inject noise in the chain without affecting the energy transport, but nonetheless causing decoherence. 
As is well known, this mechanism turns an \emph{a priori} ballistic transport, typical of harmonic lattices, into a diffusive one \cite{Bolsterli,Self_Classical_Bonetto_2004,Self_Classical_Pereira_2004,Self_Classical_Delfini_2006,Quantum_Self_cons_Dhar_2006}.
Self-consistent baths lead to a NESS current similar to that of dephasing baths, as studied for instance in  \cite{Znidaric_2010,Asadian,Asadian2013_QHO,Nicacio_2015}. 
However, they have the advantage that they lead to a Gaussian NESS (which is not the case for dephasing). Using our results, we are then able to split the total entropy production rate into a contribution stemming from the physical reservoirs at the boundaries and a second one, solely due to the self-consistent baths. We observe that each contribution scales polynomially with a distinct exponent. 
This then allows us to unambiguously associate the later with the \textit{entropic cost} of maintaining a diffusive profile in the NESS. 

This paper is structured as follows: We begin by describing our weakly interacting harmonic lattice systems and the phase-space representation of the underlying (local) master equation in Sec.~\ref{sec:QuantumPhaseSpaceMethods}. We then introduce the central object of our analysis---the Wigner entropy production---and carry out its explicit calculation and discuss the rich physics underlying its neat decomposition into elementary contributions (cf. Sec.~\ref{sec:Wigner Entropy Production Rate}). Finally,  Sec.~\ref{sec:chain}  discusses the application to estimating the entropic cost of diffusivity. 
Conclusions are summarized in Sec.~\ref{sec:conc} and, in order to make this paper self-contained, we also include a series of Appendices (\ref{app:cm}--\ref{app:separation_pis}) providing further details on the calculations outlined in Sec.~\ref{sec:chain}.

\section{Quantum phase-space methods}
\label{sec:QuantumPhaseSpaceMethods}

\subsection{The model and the quantum master equation}

We consider here a linear network of $ L $ harmonic oscillators with Hamiltonian $ \op{H} \coloneqq \op{H}_0 + \op{H}_\text{int} $, where the local part is simply $ \op{H}_0 \coloneqq \sum\nolimits_{k=1}^L H_{kk}\,\op{a}_k^\dagger\op{a}_k $ and $ \op{H}_\text{int} = \sum_{k\neq\ell} H_{k,\ell}\,\op{a}_k^\dagger \op{a}_\ell $ stands for the (linear) inter-node couplings. Here, $ \op{a}_i $ denotes the bosonic annihilation operator of the $i$-th mode and, in  what follows, $ \hbar = k_B = 1 $. Otherwise, we impose \textit{no restrictions} on the structure of the network, shaped by the non-zero off-diagonal matrix elements $ H_{\ell,k} = H_{k,\ell}^* $ ($ H_{kk} = \omega_k $).

If we let each mode interact \textit{weakly} with a local heat bath, the resulting dissipative dynamics can generally be described by a Gorini, Kossakowski, Lindblad and Sudarshan (GKLS) quantum master equation of the form \cite{Breuer_book}
\begin{equation}
\dfrac{\ud \op{\rho}}{\ud t} = \mathcal{U}(\op{\rho}) +  \sum\nolimits_{k=1}^L \mathcal{L}_k(\op{\rho}) + \pazocal{O}(\gamma^2).
\label{global_master_eqn}
\end{equation}
Here, $ \mathcal{U}(\op{\rho}) \coloneqq -i [\op{H}, \op{\rho}] $ and $ \mathcal{L}_k $ stands for a dissipation super-operator in the standard GKLS form \cite{Gorini_1976,lindblad1976generators}. Moreover,  $ \gamma \coloneqq \max \{\gamma_1,\cdots,\gamma_L\} $ equals the largest of the node--bath coupling strengths $ \gamma_k $, i.e., it carries the order of magnitude of the dissipative interactions. Individually, each super-operator $ \mathcal{L}_k $ is $ \pazocal{O}(\gamma_k) $. In general, any of the ``dissipators'' $ \mathcal{L}_k $ can act \textit{globally} on all nodes of the network. One may, however, make the additional assumption of weak internal coupling between the nodes, which leads to the \textit{local} GKLS quantum master equation \cite{trushechkin2016perturbative}
\begin{equation}
\dfrac{\ud \op{\rho}}{\ud t} = -i [\op{H}, \op{\rho}] +  \sum\nolimits_{k=1}^L \mathcal{D}_k(\op{\rho}) + \pazocal{O}(\lambda\,\gamma).
\label{master_eqn}
\end{equation}
where the local dissipators are given by
\begin{multline}
\mathcal{D}_k(\op{\rho}) =  \gamma_k (n_k +1)\left(\op{a}_k \op{\rho} \op{a}_k^\dagger - \frac{1}{2}\{\op{a}_k^\dagger \op{a}_k,\op{\rho}\}_+\right) \\
+ \gamma_k n_k \left(\op{a}_k^\dagger \op{\rho} \op{a}_k - \frac{1}{2}\{\op{a}_k \op{a}_k^\dagger,\op{\rho}\}_+ \right),
\label{Dissipator}	
\end{multline}
with $ n_k \coloneqq (e^{\omega_k/T_k}-1)^{-1} $ being the Bose--Einstein thermal occupation of at temperature $ T_k $ and frequency $ \omega_k $, and where $ \{\cdot,\cdot\}_+ $ stands for anti-commutator.

Care must be taken when using Eq.~\eqref{master_eqn} to describe a multipartite open quantum system, since it is well known that going beyond its range of validity might lead to thermodynamic inconsistencies \cite{levy2014local,stockburger2016thermodynamic,trushechkin2016perturbative}. The parameter range in which the many approximations underlying \textit{both} global and local master equations are satisfied have been critically (and extensively) discussed in the literature \cite{Rivas_10,Wichterich_07,Current_drag_09,Purkayastha_16,Luis_Adesso_LME_GME_17,Brunner_2017,seah2018refrigeration,kolodynski2018adding,Plenio_2018,naseem2018thermodynamic,Stella_2018}. Interestingly, local master equations also arise naturally when considering certain repeated-interaction models \cite{Current_drag_09,Barra2015,Stella_2018}. This picture enables a thermodynamically consistent bookkeeping of \textit{all} energy exchanges occurring in dissipative processes \textit{exactly} described by Eq.~\eqref{master_eqn} \cite{barra2017smallest,de2018reconciliation}. In this paper, we remain on the safe side by working well within the range of applicability of Eq.~\eqref{master_eqn}, understood as a mere perturbative expansion of Eq.~\eqref{global_master_eqn} to lowest order in the internal couplings.

\subsection{The Fokker--Planck equation}

We now move from Hilbert space to the quantum phase space. Introducing the displacement operator $ \op{D}(\bar{\beta}) \coloneqq \exp{\big(\sum\nolimits_{k=1}^{L}\beta_k \op{a}_k^\dagger - \beta_k^* \op{a}_k\big)} $, the Wigner function of our $L$-mode Gaussian state $\op{\rho}$ can be written as \cite{Quatum_noise_gardiner}
\begin{equation}
W(\bar{\alpha}) \coloneqq \dfrac{1}{\pi^{2L}} \int \ud \bar{\beta} ~e^{-\sum_{k} (\beta_k \alpha_k^* - \beta_k^*\alpha_k)} \tr\left\{\,\op{\rho}\, \op{D}(\bar{\beta})  \right\},
\label{Wigner_definition}
\end{equation}
where $ \bar{\alpha} $ and $ \bar{\beta} $ are $2L$-dimensional complex vectors containing phase-space variables, i.e., $\bar{\alpha} \coloneqq (\alpha_1, \alpha_1^*, \ldots, \alpha_L, \alpha_L^*)^\top $. The reason for choosing the Wigner representation lies in the Gaussianity of $ \op \rho $, which brings at least three advantages. First, it is always strictly positive, which allows us to identify the Wigner function as a quasi-probability distribution. The second, it will depend only on the first and second order moments, which greatly simplifies the analysis (cf. Appendix \ref{app:cm} for details). Finally, the corresponding Wigner entropy, to be introduced below, can be related to the R\'enyi-2 entropy and also satisfies the strong subadditivity inequality \cite{Adesso_12_WignerEntropy}, hence giving it a more physical appeal. 

Using standard correspondence tables (see, e.g., \cite{Quatum_noise_gardiner}) one can turn Eq.~\eqref{master_eqn} into the following Fokker--Planck equation
\begin{equation}
	\partial_t W = \pazocal{U}(W) + \sum\nolimits_{k=1}^L \pazocal{D}_k(W),
	\label{fokker_planck_eq}
\end{equation}
where the phase-space super-operator $\pazocal{U}(W)$ represents the unitary part of Eq.~\eqref{master_eqn}, while $ \pazocal{D}_k(W)$ stands for the dissipative contributions. In particular, the latter may be written as divergences in the complex plane; namely
\begin{equation}\label{eq:dissipator_Wigner}
\pazocal{D}_k(W) = \partial_k \pazocal{J}_k(W) + \partial^*_{k} \pazocal{J}_k^*(W),
\end{equation}
where $\partial_k  \coloneqq \partial/\partial \alpha_k$ ($\partial_k^* \coloneqq \partial/\partial \alpha_k^*$) and 
\begin{equation}\label{currents}
\pazocal{J}_k(W) = \frac{\gamma_k}{2} \left[\alpha_k W + \left(n_k+1/2\right) \partial_k^* W \right].
\end{equation}

Eq.~\eqref{fokker_planck_eq} can be interpreted as a continuity equation for $ W $ with $ \pazocal{J}_k(W)$ playing the role of irreversible quasi-probability currents in phase space. In fact, these currents are identically zero if and only if each oscillator is in local thermal equilibrium~\footnote{This follows directly from the fact that $ \op{\rho}_\text{eq} = \bigotimes_k\,\op{\rho}_\text{eq}^{(k)} $, with $ \op{\rho}_\text{eq}^{(k)} = \exp{\big(-\omega_k\,\op{a}_k^\dagger\op{a}_k/T_k\big)} $, is the unique fixed point of the dissipative part of the local master equation \eqref{master_eqn}. Notice, however, that $ \ud\op{\rho}_\text{eq}/\ud t \neq 0 $; not even within the range of validity of the local approximation. Specifically, $ \ud\op{\rho}_\text{eq}/\ud t = -i[\op{H},\op{\rho}_\text{eq}] = -i[\op{H}_\text{int},\op{\rho}_\text{eq}] = \pazocal{O}(\lambda) \gg \pazocal{O}(\lambda\,\gamma)$.}; i.e., $ \pazocal{J}_k(W_\text{eq}) = 0 ~\forall k $ with
\begin{equation}\label{Wigner_eq}
W_\text{eq} = \prod\nolimits_{k=1}^L W_\text{eq}^{(k)}, \qquad
W_\text{eq}^{(k)} = \dfrac{e^{-\vert\a_k\vert^2/(n_k + 1/2)}}{\pi\,(n_k + 1/2)}. 
\end{equation}
Since it is \textit{only} in $ W_\text{eq} $ that all the individual currents $ \pazocal{J}_k(W) $ vanish exactly, we shall adopt it as the reference state for quantifying entropy-production rates and fluxes \cite{Santos_Wigner_2017}. In this limited sense, $ W_\text{eq} $ would correspond to the thermal equilibrium state in the standard formulation.

Similarly, the unitary part $\pazocal{U} $ in Eq.~\eqref{fokker_planck_eq} can be cast as 
\begin{equation}\label{unitary_currents}
\pazocal{U}(W) = \sum\nolimits_{k=1}^{L}  \left[\pt_{k} \pazocal{A}_{k}(W) +  \partial_{k}^*\pazocal{A}_{k}^*(W) \right],
\end{equation}
where, for our choice of Hamiltonian, the \textit{reversible} quasi-probability currents $\pazocal{A}_k(W)$ are simply
\begin{equation}
\pazocal{A}_k(W) = i \sum\nolimits_{\ell=1}^{L}  H_{k \ell}\,\a_\ell\,W.
\end{equation}

\section{Wigner entropy-production rate}
\label{sec:Wigner Entropy Production Rate}

\subsection{Individual dissipative contributions}

We shall now decompose $ \ud S/\ud t $ as in Eq.~\eqref{dSdt_Pi_PHI} into an entropy-production rate and the entropy flux. We base our analysis in the \textit{Wigner entropy} defined as
\begin{equation}\label{Wigner_entropy}
S(W) \coloneqq -\int \ud \bar{\alpha} \; W(\bar{\alpha}) \ln{W(\bar{\alpha})}.
\end{equation}
As shown in Refs.~\cite{Santos_Wigner_2017,Brunelli_2016,Santos_spin_2018,Santos_BathView_2018}, for Gaussian systems this is entirely equivalent to the standard approach based on von Neumann entropy. Besides, the Wigner entropy offers several advantages, as we shall see below.

Differentiating Eq.~\eqref{Wigner_entropy} with respect to time and using Eq.~\eqref{fokker_planck_eq} yields 
\begin{equation}\label{dSdt_step1}
\dfrac{\ud S}{\ud t} = -\int \ud\bar{\alpha}\,\left[ \pazocal{U}(W) + \sum\nolimits_k \pazocal{D}_k(W) \right]\ln{W}.
\end{equation}
Adding and subtracting $ \sum\nolimits_k\int\ud\bar{\alpha}\,\pazocal{D}_k(W)\ln{W_\text{eq}} $, so as to introduce our fiducial state $ W_\text{eq} $, gives
\begin{align}\label{dSdt_step2}
\dfrac{\ud S}{\ud t} &= -\sum\nolimits_k\int \ud\bar{\alpha}\, \pazocal{D}_k(W)\ln{(W/W_\text{eq})}\nonumber\\
&-\sum\nolimits_k\int \ud\bar{\alpha}\, \pazocal{D}_k(W)\ln{W_\text{eq}} \coloneqq \Pi - \Phi, 
\end{align}
where we have used the fact that the unitary dynamics $ \pazocal{U} $ does not change the Wigner entropy. 
The identification of the two terms in this equation as an entropy production rate and an entropy flux rate will be better justified once they are evaluated explicitly, as we shall now do.

Inserting Eq.~\eqref{eq:dissipator_Wigner} in $ \Pi $, we can integrate by parts in each variable. Noticing that the corresponding boundary terms vanish due to the Gaussianity of $ W $ and $ W_\text{eq} $, we get
\begin{equation}
\Pi = \sum\nolimits_k \int \ud \bar{\alpha} \left[ \pazocal{J}_k(W)\,\partial_k + \pazocal{J}_k^*(W)\,\partial_{k}^* \right]\,\ln{(W/W_\text{eq})}.
\end{equation}
Next, we rewrite the currents $ \pazocal{J}_k(W)$ in Eq.~\eqref{currents} as 
\begin{equation}\label{currents_log}
\pazocal{J}_k(W) = \dfrac{\gamma_k}{2} (n_k + 1/2)\,W\,\partial_{k}^* \ln{(W/W_\text{eq})},
\end{equation}
which allows us to express $\partial_{k}^* \ln{(W/W_\text{eq})}$ in terms of $\pazocal{J}_k(W)$ and $ W $. This leads to the decomposition $ \Pi = \sum\nolimits_k\,\Pi_k $, where
\begin{equation}\label{eq:Pi_k}
\Pi_k = \dfrac{4}{\gamma_k (n_k+ 1/2)} \int \ud\bar{\alpha}\,\dfrac{\left\vert \pazocal{J}_k(W) \right\vert^2}{W}.  
\end{equation}
As shown in Appendix~\ref{app:Pi_CM}, further manipulations lead to a compact expression in terms of the covariances of the system. Similarly, the entropy flux decomposes as $ \Phi = \sum\nolimits_k\Phi_k $ with
\begin{equation}
\label{Phi_k}
\Phi_k =  -\int \ud\bar{\alpha}\,\left[ \pazocal{J}_k(W)\,\partial_{k} \ln{W_\text{eq}} + \pazocal{J}^*_k(W)\,\partial_{k}^* \ln{ W_\text{eq}} \right] .
\end{equation}
Making use of the identity
\begin{equation}\label{ave_Wigner}
\int \ud\bar{\alpha}\,\alpha_k^* \alpha_\ell W  = \langle \op{a}_k^\dagger \op{a}_\ell \rangle + \frac12\delta_{k,\ell},
\end{equation}
we finally get the compact expression
\begin{equation}\label{Phi_k_2}
\Phi_k = \frac{\gamma_k}{n_k+ 1/2} \left(\langle \op{a}_k^\dagger \op{a}_k \rangle - n_k\right).
\end{equation}

Eqs.~\eqref{eq:Pi_k} and \eqref{Phi_k_2} constitute our first main result. Their identification as entropy-production rate and entropy flux, respectively, is  based on several supporting arguments and has been extensively debated in the past, both in the quantum \cite{Santos_Wigner_2017} and classical \cite{Classical_th_review_Seifert_2012,Fok_Planck_class_Tome_2010} contexts. First and foremost, $ \Pi $ is clearly non-negative and zero if and only if the currents themselves vanish, which only occurs in the reference state $ W_\text{eq} $. Second, the proposed entropy-production rate is an even function of the irreversible currents whereas the entropy flux is odd, in agreement with other studies based on Fokker--Planck equations \cite{Fok_Planck_class_Tome_2010}. Finally and most remarkably, it can be shown that, within the framework of stochastic trajectories, this expression satisfies integral fluctuation theorems \cite{Santos_Wigner_2017,Fok_Planck_class_Spinney_2012}. 

Rather intuitively, we can also see that Eq.~\eqref{Phi_k_2} is proportional to the difference between the actual occupation of the $k$-th mode and the equilibrium occupation at the temperature $T_k$ of the local bath. Namely, if the mode ``looks hotter'' than its local environment, one has $ \Phi_k > 0 $, i.e., entropy flows from the system into the bath, as expected. Conversely, if $ \langle \op{a}_k^\dagger \op{a}_k \rangle < n_k$, $\Phi_k < 0 $ and entropy flows into the system. Note that $ \Phi_k $ is, therefore, an inherently \textit{observable} quantity \cite{Santos_Wigner_2017}.

For us, the most relevant feature of Eqs.~\eqref{eq:Pi_k} and \eqref{Phi_k_2} is precisely that they provide the individual contribution of each dissipation channel to the the total entropy-production rate and entropy flux. In particular, from a classical viewpoint \cite{Classical_th_review_Seifert_2012} one may understand Eq.~\eqref{eq:Pi_k} as an average of the ``phase-space velocity'' $\pazocal{J}_k(W)/W$ of each individual quasi-probability current $\pazocal{J}_k(W)$. Hence, the entropy production can be thought-of as weighted average of such mean velocities. 

\subsection{Role of the unitary dynamics in maintaining a NESS}

Let us introduce the \textit{Wigner relative entropy} (or Kullback--Leibler divergence), defined as
\begin{equation}
\infdiv{W}{W_\text{eq}} \coloneqq \int \ud \bar{\alpha}\, W \ln{(W/W_\text{eq})},
\end{equation}
and calculate its rate of change using Eq.~\eqref{fokker_planck_eq}. This yields
\begin{align}\label{derivative_relat_entropy}
\frac{\ud}{\ud t} \infdiv{W}{W_\text{eq}}  &= \int \ud\bar{\alpha}\,\pazocal{U}(W) \ln{(W/W_\text{eq})} \nonumber \\
 &+  \sum\nolimits_{k=1}^L \int \ud\bar{\alpha}\,\pazocal{D}_k(W) \ln{(W/W_\text{eq})} \nonumber \\
 &=  -\int \ud\bar{\alpha}\,\pazocal{U}(W) \ln{W_\text{eq}} - \Pi,
\end{align}
where we have used Eq.~\eqref{dSdt_step2}. In the long-time limit, the stationarity condition $\ud\infdiv{W_\text{NESS}}{W_\text{eq}}/\ud t = 0$ entails 
\begin{equation}
\label{Pi_unitary}
\Pi(t\rightarrow\infty)\coloneqq\Pi_\text{NESS} = -\int \ud\bar{\alpha}\,\pazocal{U}(W_\text{NESS}) \ln{W_\text{eq}}.
\end{equation}
This is our second main result: it illustrates the essential role of the unitary internal dynamics in sustaining stationary heat currents across the network which, in turn, translates into a finite rate of steady-state entropy production.


Inserting Eq.~(\ref{unitary_currents}) for the unitary currents  and using \eqref{ave_Wigner} we also find that the last term in Eq.~(\ref{Pi_unitary}) may be written as 
\begin{align}
\Pi_\text{NESS} &= - i \sum\nolimits_{k\neq\ell} \frac{1}{n_k+1/2} \left(H_{k,\ell}\,\langle a_k^\dagger a_\ell \rangle - H_{\ell,k}\,\langle a_\ell^\dagger a_k \rangle\right) \nonumber\\
&= \sum\nolimits_{k\neq\ell} \frac{2}{n_k+1/2}\,\text{Im}\,\lbrace H_{k,\ell}\,\langle a_k^\dagger a_\ell \rangle\rbrace.
\label{Pi_unitary_2} 
\end{align}
We point to the fact that $ \Pi_\text{NESS} $ in Eq.~\eqref{Pi_unitary_2} contains exclusively off-diagonal elements $ H_{k\neq\ell} $ and none of the local components $ H_{kk}\,\op{a}_k^\dagger\op{a}_k $; it is thus entirely due to the node--node couplings. Introducing the \textit{energy currents} $ j_{k,\ell} $ defined by \cite{Current_drag_09} 
\begin{subequations}
\begin{align}\label{current_sites}
\frac{\ud \langle \op{a}_k^\dagger \op{a}_k \rangle}{\ud t} &= i\,\langle [\op{H},\,\op{a}_k^\dagger \op{a}_k]\rangle \coloneqq \sum\nolimits_{\ell \neq k} j_{k,\ell},\\
j_{k,\ell} &= -i\,H_{k,\ell} \left(\langle \op{a}_k^\dagger \op{a}_\ell \rangle - \langle \op{a}_\ell^\dagger \op{a}_k \rangle\right) = -j_{\ell,k}. \label{eq:local_current_definition}
\end{align}
\end{subequations}
allows to cast Eq.~(\ref{Pi_unitary_2}) in the symmetric form 
\begin{equation}\label{Pi_unitary_3}
\Pi_\text{NESS} = \frac{1}{2}\sum\nolimits_{k\neq\ell}\,j_{k,\ell}\,\left(\frac{1}{n_k + 1/2} - \frac{1}{n_\ell + 1/2} \right).
\end{equation}

This result can be connected with Onsager's theory of irreversible thermodynamics \cite{Callen,Onsager_class_Onsager_1931}. 
Within this framework, the entropy production is defined as the product of `fluxes' and `affinities' (also called `generalized forces'). 
For instance, the current of energy is related to the affinity $1/T$ so that, in a classical scenario, the Onsager entropy production between two bodies kept at temperatures $T_A$ and $T_B$ is given by $ \Pi = j_{AB}\,\big(1/T_A - 1/T_B\big)$, where $j_{AB}$ stands for the energy current from $B$ to $A$. We see that Eq.~\eqref{Pi_unitary_3} has the exact same mathematical structure, which is yet another consistency test for our analysis. Moreover, notice that, due to the fact that we are using Wigner entropies, the thermodynamic affinity related to the current is not the inverse temperature, but rather the inverse Bose--Einstein occupation $ n + 1/2 $. For high temperatures $ n + 1/2 \propto T$ so that both frameworks coincide. However, our results hold true even in the limit of vanishingly low temperature\footnote{Always provided that the Markov approximation behind the master equation \eqref{master_eqn} continues to hold. This is often seen as a high-temperature limit.}. We also remark that Onsager's formula is valid only close to equilibrium (linear response theory), whereas Eq.~\eqref{Pi_unitary_3} holds true for arbitrary non-equilibrium states. This is a consequence of the Gaussianity of the problem. 

Finally note that, as $ \Pi_\text{NESS} \geq 0 $, Eq.~\eqref{Pi_unitary_3} implies that $ j_{k,\ell} $ has the same sign as $ 1/(n_k + 1/2) - 1/(n_\ell + 1/2) $; i.e., when node $k$ is hotter than node $\ell$, the local current $ j_{k,\ell} $ flows from $ k $ to $ \ell $.

\section{\label{sec:chain} Application: The entropic cost of diffusivity}

\begin{figure*}[t!]
	\centering
	\includegraphics[width=0.8\textwidth]{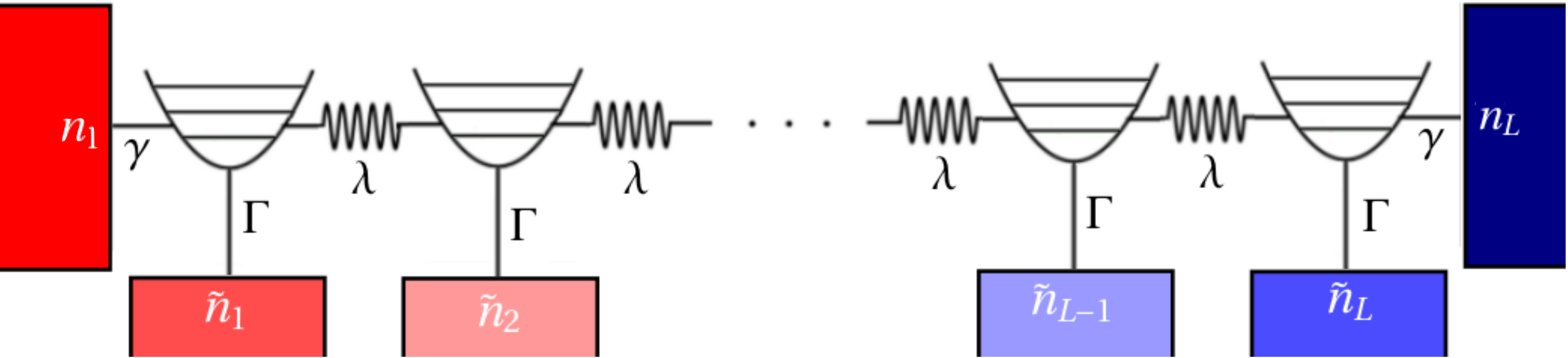}
	\caption{(Color online). Schematic diagram of our model: A one-dimensional chain comprised of $L$ identical harmonic oscillators of frequency $ \omega $ linearly coupled via nearest-neighbor interactions of (weak) strength $ \lambda $, is put in contact with two reservoirs kept at temperatures $ T_1 $ and $ T_L $, respectively or, equivalently, occupations $n_1$ and $n_L$ at frequency $ \omega $. To induce diffusivity, the system is augmented with $L$ self-consistent reservoirs (one for each site of the chain). The occupations for these reservoirs ($\tilde{n}_k$) is chosen so that, in the steady state, no energy is exchanged with them and the heat current flows only between the physical reservoirs.
	}
	\label{fig:chain_draw}
\end{figure*}

In order to illustrate the usefulness of this framework, we now apply it to a typical quantum transport problem. In particular, we show how to exploit it to quantify the irreversibility associated with sustaining \textit{diffusive} heat transfer through a chain of harmonic oscillators. 

Before describing the model further, let us briefly discuss the notions of `ballistic' and `diffusive' transport. Consider the NESS established when a system (of ``length'' $L$) is placed between two reservoirs at different temperatures. This exhibits a stationary heat current which, for small temperature gradients $\Delta T$, may be fitted by $j = \kappa\,\Delta T/ L^{\alpha}$, where $\alpha$ is an exponent that characterizes the phenomenological heat transport law. Whenever $ \alpha = 1 $ (i.e., Fourier's heat conduction law) one speaks of \textit{diffusive} behavior. However, low-dimensional---and usually integrable---quantum and classical systems can display a \textit{ballistic} heat flow \cite{Rieder_HarmonicChain_Classical_Ballistic,energy_cons_noise_Landi_2013,Dhar_08}, characterized by $\alpha = 0$; that is, a heat current that is independent of the system size. 


Here, we aim at understanding how the entropy-production rate in the steady state relates to the heat transfer law; specifically, considering the divide between diffusive and ballistic heat conduction. To do so, we must adopt a model allowing to switch between these two regimes. In particular, our one-dimensional harmonic chain coupled to two different reservoirs at each end would result in a ballistic NESS, provided that its dissipative dynamics is well described by Eq.~\eqref{master_eqn}. 
Obtaining diffusive behavior in general requires complex interactions, which are seldom treatable analytically. 
An alternative, widely used in the literature on quantum systems, is to use dephasing baths acting locally on each site \cite{Znidaric_2010,Asadian,Asadian2013_QHO}. 
These baths introduce some noise while not compromising the energy balance. 
As a consequence, they always lead to diffusive transport. 

The NESS for a dephasing model, however, is not Gaussian, so that the results presented here would not apply. 
Instead, we fix this problem by using the notion of `self consistent reservoirs'  \cite{Bolsterli,Znidaric_2010,Asadian,Asadian2013_QHO,Nicacio_2015} (also called B\"uttiker probes in the condensed matter literature). As the name suggests, self-consistent reservoirs are baths that act on all sites, but whose temperature are chosen self-consistently so that, in the steady state, they do not exchange any energy with the system, but only inject noise to cause decoherence \cite{Bolsterli}. Consequently, heat only flows between the two physical reservoirs at the ends of the chain (see Fig.~\ref{fig:chain_draw}).
As shown in Appendix~\ref{app:CM_chain_NESS}, the resulting equation of motion for the covariances is the same for  self-consistent and dephasing models although the corresponding NESSs are different (i.e., they only agree up to the second-order moments). 


\subsection{\label{ssec:model}The model and its stationary solution}

In what follows, we consider $L$ resonant harmonic oscillators in a one-dimensional lattice interacting weakly via the linear Hamiltonian 
\begin{equation}
\op{H} = \omega \sum\nolimits_{k=1}^L \op{a}_k^\dagger \op{a}_k + i \lambda \sum\nolimits_{k=1}^{L-1} \left(\op{a}_{k}^\dagger \op{a}_{k+1} - \op{a}_{k+1}^\dagger \op{a}_k\right).
\label{Hamiltonian1b}
\end{equation}
As described above, the system is coupled to $ L + 2 $ baths, out of which two are physical and the rest, auxiliary self-consistent reservoirs. The master equation \eqref{master_eqn} then reads 
\begin{equation}\label{M_model}
\dfrac{\ud \op{\rho}}{\ud t} = -i [\op{H}, \op{\rho}] +  \mathcal{D}_1(\op{\rho}) + \mathcal{D}_L(\op{\rho}) + \sum\nolimits_{k=1}^L \tilde{\mathcal{D}}_k(\op{\rho}).
\end{equation}
Here all dissipators have the GKLS structure of Eq.~\eqref{Dissipator}. Those of the physical baths have parameters $\gamma_1 = \gamma_L = \gamma$ and the corresponding occupations at frequency $\omega$ are $n_1$ and $n_L$. On the other hand, the auxiliary baths all have dissipation rate $\tilde{\gamma}_k = \Gamma$ and occupations $\tilde{n}_k$, calculated self-consistently so as to satisfy $\tilde{n}_k = \tr\{\op{a}_k^\dagger \op{a}_k\,\op{\rho}_\text{NESS}\}$. All objects denoted with tilde refer to the auxiliary baths. In particular, notice the difference between $ \tilde{\mathcal{D}}_{1,L}(\op{\rho}) $, or $ \tilde{n}_{1,L} $, and $ \mathcal{D}_{1,L}(\op{\rho}) $ and $ n_{1,L} $.

The stationary state of Eq.~\eqref{M_model} (i.e., $ \op{\rho}_\text{NESS} $) can be found analytically with the methods developed in Refs.~\cite{Znidaric_2010,Asadian,Asadian2013_QHO,Nicacio_2015}. For completeness, we give details of the calculation in Appendix~\ref{app:CM_chain_NESS}. We find that the only non-zero stationary covariances in our chain are 
\begin{subequations}
	\label{eq:stationary_covariances}
\begin{multline}
\langle \op{a}_k^\dagger \op{a}_k \rangle = \frac{n_1 + n_L}{2} + \frac{n_1-n_L}{2} \\
\times\frac{\Gamma\,\gamma\,(L-2k+1) + \gamma^2 \left(\delta_{1,k}  -\delta_{k,L}\right)}{4\lambda^2 + \gamma^2 + \gamma\,\Gamma(L-1)},
 \label{Ckk_example}
\end{multline}
\begin{equation}
 \langle \op{a}_k^\dagger \op{a}_{k+1} \rangle = \dfrac{\gamma\,\lambda\,(n_L-n_1)}{4\lambda^2+\gamma^2+ \gamma\,\Gamma\,(L-1)}.
 \label{corr_example}
\end{equation}
\end{subequations}

\begin{figure}
	\centering
	\includegraphics[width=0.22\textwidth]{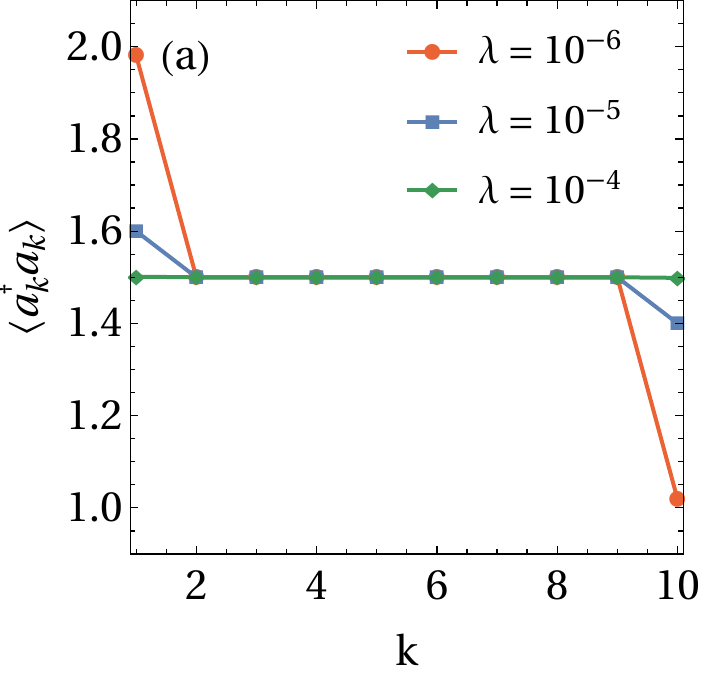}\quad
	\includegraphics[width=0.22\textwidth]{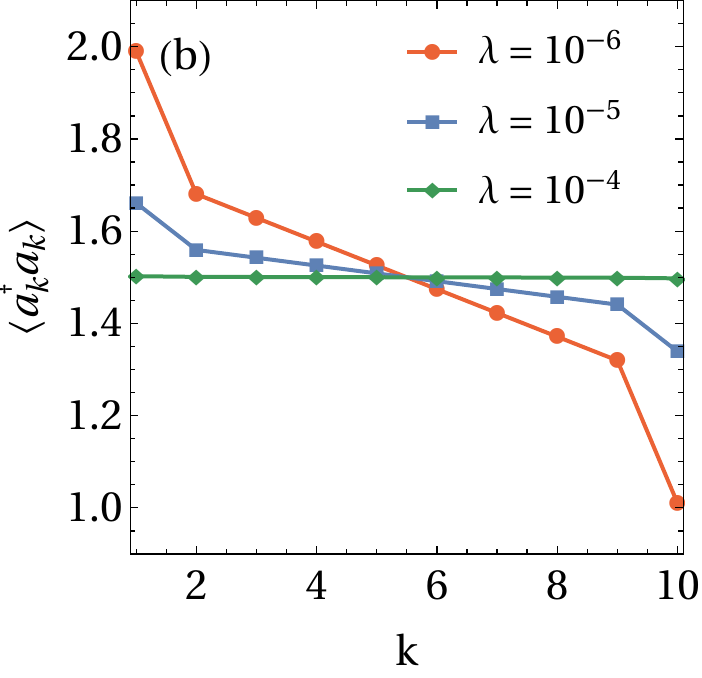}\\
	\includegraphics[width=0.22\textwidth]{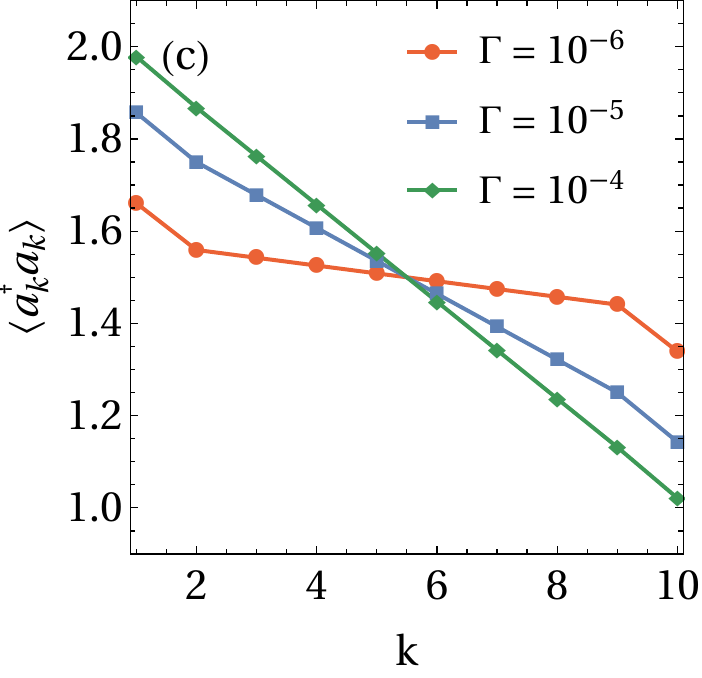}\quad
	\includegraphics[width=0.22\textwidth]{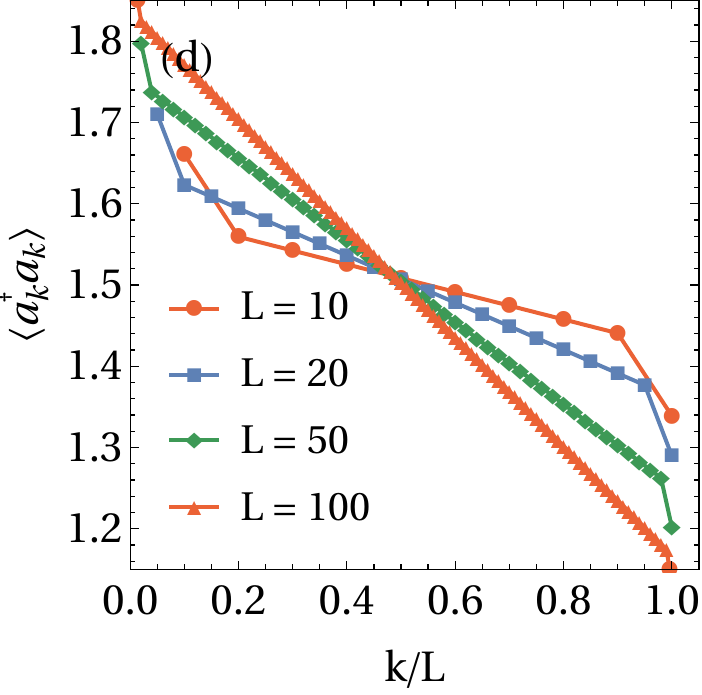}
	\caption{\label{fig:profile}
		(Color online). Occupation profile $\langle \op{a}_k^\dagger \op{a}_k\rangle$ in the NESS $ \op{\rho}_\text{NESS} $ of Eq.~\eqref{M_model} as a function of the location $ k $ of the site on the chain for various parameters. 
		\textbf{(a)} Ballistic profile for $\Gamma = 0$ and varying $\lambda$.
		\textbf{(b)} Diffusive profile for $\Gamma = 10^{-6}$. 
		\textbf{(c)} Same as (b) but for different values of $\Gamma$, with fixed $\lambda = 10^{-5}$. 
		\textbf{(d)} Profile for different chain sizes $L$, with fixed $\lambda = 10^{-5}$ and $\Gamma = 10^{-6}$. 
		In all panels $\gamma = 10^{-5}$. In (a)--(c) $L = 10$.
	}
\end{figure}

Eqs.~\eqref{eq:stationary_covariances} have a rich physical interpretation. First, setting $\Gamma = 0$ amounts to decoupling the self-consistent reservoirs from the chain. As already anticipated, in that case the occupations $\langle \op{a}_k^\dagger \op{a}_k \rangle$ become independent of the index $ k $ (except for $ k =1 $ and $ k= L $), which is the hallmark of a ballistic heat transfer law [see Fig.~\ref{fig:profile}(a)]. Conversely, if we choose $\Gamma > 0$ and sufficiently large $L$, the occupations converge to a linear profile, which is expected for diffusive heat conduction in one dimension. This is illustrated in Figs.~\ref{fig:profile}(b)--(d).

Furthermore, combining Eqs.~\eqref{eq:local_current_definition} and \eqref{corr_example} we can compute the current between neighboring sites
\begin{equation}\label{currents_NESS_example}
j_{k,k+1} = \frac{2 \lambda^2\,\gamma \left(n_L - n_1\right)}{4\lambda^2+\gamma^2+\gamma\,\Gamma(L-1)} \coloneqq j,
\end{equation}
which are all independent of $ k $. Here we see once again that if $\Gamma = 0$, the energy current becomes independent of $L$, implying a ballistic transport. On the other hand, $\Gamma > 0$ yields a diffusive heat current, scaling as $\sim 1/L$. We remark that discussing the divide between ballistic and diffusive conduction is only meaningful in the thermodynamic limit (i.e., $ L \gg 1 $), where our model suffers an abrupt transition from a ballistic profile to a diffusive one as soon as $\Gamma$ becomes non-zero. 

\subsection{Entropy-production rate in the NESS}

\begin{figure}[b!]
	\centering
	\includegraphics[width=0.45\textwidth]{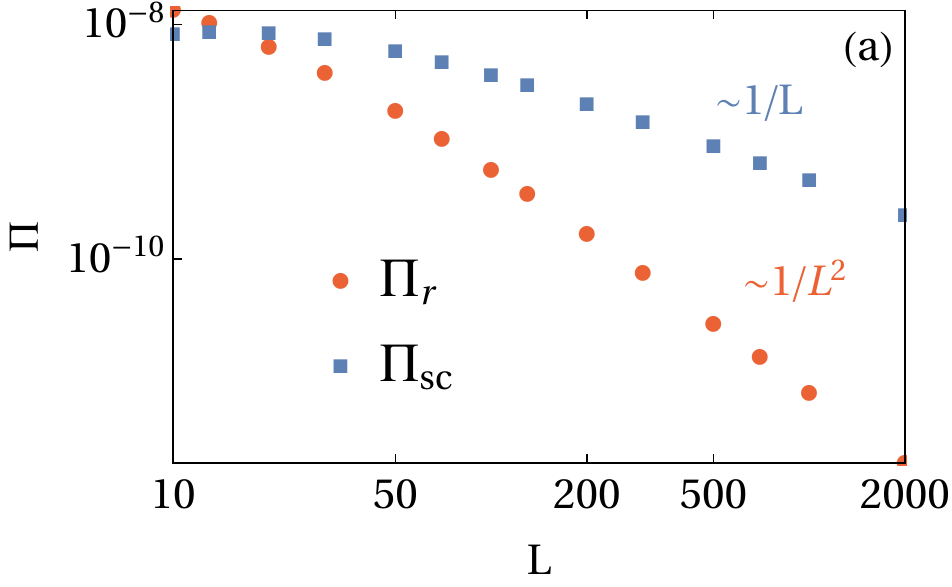}
	\caption{\label{fig:Pi_L}
		(Color online). Individual contributions to the total stationary entropy-production rate from the real (red dots) and the self-consistent reservoirs (blue squares), as a function of the system size $L$. Here, $\gamma = 10^{-6}$, $\Gamma = 10^{-7}$, $\lambda = 3\times10^{-7}$, $n_1 = 1$, and $n_L = 2$. 
	}
\end{figure}

The \textit{total} (Wigner) entropy-production rate can be readily computed from Eqs.~\eqref{Pi_unitary_3} and \eqref{currents_NESS_example}. Since the currents $ j_{k,k+1} $ are translationally invariant, all inner terms in the sum of Eq.~\eqref{Pi_unitary_3} vanish and we are left with 
\begin{equation}\label{Pi_ness_example}
\Pi_\text{NESS} = \frac{2 \lambda^2\,\gamma\,\left(n_L-n_1\right)}{4\lambda^2+\gamma^2+\gamma\,\Gamma\,(L-1)}\,\left( \frac{1}{n_1 + 1/2} - \frac{1}{n_L + 1/2}\right).
\end{equation}
Note that this is always non-negative and zero only if $ n_1 = n_L $, in which case there is no current flowing through the chain. We also remark that neither $ j $ nor $ \Pi_\text{NESS} $ depend on $\tilde{n}_k$. This is an important consistency check, as our goal is precisely to ensure that the auxiliary reservoirs do not modify the rate of heat flow across the chain. Notice as well that $ \Pi_\text{NESS} $ scales in size just like the rate of heat exchange $ j $.

Importantly, exploiting Eq.~\eqref{eq:Pi_k} we are also able to separate the contributions of each dissipation channel to the stationary entropy-production rate. More specifically, we can split $ \Pi_\text{NESS} $ into one component stemming from the real reservoirs at the boundaries and another one coming from the auxiliary self-consistent reservoirs; that is, $\Pi_\text{NESS} \coloneqq \Pi_\text{r} + \Pi_\text{sc}$. Crucially, due to our approach to emulate the anharmonicity leading to the diffusive profile, we can meaningfully identify the self-consistent contribution $ \Pi_\text{sc} $ with the entropic cost of steady-state diffusivity. Specifically, one has
\begin{equation}
\label{Pi_r_example}
\Pi_\text{r} =  \sum\nolimits_{k=1,L}\frac{4}{\gamma\,\left(n_k+ 1/2\right)} \int \ud \bar{\alpha} \frac{\vert \pazocal{J}_k(W_\text{NESS}) \vert^2}{W_\text{NESS}},
\end{equation}
while the entropy production from the self-consistent reservoirs reads 
\begin{subequations}
\label{Pi_sc_example}
\begin{align}
\Pi_\text{sc} &=  \sum\nolimits_{k=1}^L \frac{4}{\Gamma\,\left(\langle \op{a}_k^\dagger \op{a}_k \rangle + 1/2\right)} \int \ud\bar{\alpha} \frac{\vert\tilde{\pazocal{J}}_k(W_\text{NESS})\vert^2}{W_\text{NESS}}, \label{Pi_sc_example-1}\\
\tilde{\pazocal{J}}_{k}(W_\text{NESS}) &= \frac{\Gamma}{2}\,\left[\alpha_{k} W_\text{NESS} + \big(\langle \op{a}_k^\dagger \op{a}_k \rangle + 1/2 \big)\,\partial_{k}^* W_\text{NESS} \right],
\end{align}
\end{subequations}
where we already used the fact that the occupations of the self-consistent reservoirs are $\tilde{n}_k = \langle \op{a}_k^\dagger \op{a}_k \rangle$. The sum in Eq.~\eqref{Pi_r_example} comprises the terms $ k = 1 $ and $ k = L $ only, while that of Eq.~\eqref{Pi_sc_example-1} runs over all indices $ k \in \{1,\cdots, L\} $. As discussed in Appendix~\ref{app:separation_pis}, Eqs.~\eqref{Pi_r_example} and \eqref{Pi_sc_example-1} can be readily computed from the covariances of the chain for arbitrarily large system size, owing to the Gaussianity of our NESS. 

In Fig.~\ref{fig:Pi_L} we plot $ \Pi_\text{r} $ and $ \Pi_\text{sc} $ as a function of $ L $ in a logarithmic scale. Interestingly, we see that the irreversibility associated with the physical reservoirs is dominant \textit{only} for small system size. As the chain is scaled up $ \Pi_\text{sc} $ quickly surpasses $ \Pi_\text{r} $ and adopts a distinct power-law-like decay $ \Pi_\text{sc} \sim 1/L $ for large $ L $. On the other hand---and also in the thermodynamic limit---we observe that the contribution from the real baths to the total steady-state irreversibility decays as $ \Pi_\text{r} \sim 1/L^2 $. Looking back at Eq.~\eqref{Pi_ness_example}, it is also noteworthy that, fixing the temperature gradient, the ballistic irreversible entropy-production rate $ \Pi_\text{NESS}(\gamma,\Gamma = 0) $ is larger than the diffusive one $ \Pi_\text{NESS}(\gamma,\Gamma > 0) $, meaning that $ \Pi_\text{r}(\gamma,\Gamma = 0) > \Pi_\text{r}(\gamma,\Gamma\neq 0) + \Pi_\text{sc}(\gamma,\Gamma\neq 0) $. The dependence on \textit{both} $ \gamma $ and $ \Gamma $, although not explicit in Eqs.~\eqref{Pi_r_example} and \eqref{Pi_sc_example}, arises from $ W_\text{NESS} $.

We have thus seen that combining the ease of calculation of the Wigner entropy production for Gaussian states with our simple harmonic model to mimic diffusive heat conduction, provides valuable insights into the relative weight of the individual irreversible processes in the thermodynamic limit. 

\section{\label{sec:conc}Conclusions}

We have addressed the calculation of the irreversible entropy-production rate using a quantum phase-space approach based on the Wigner entropy. In particular, we focused on networks of weakly interacting harmonic nodes with arbitrary connectivity, and coupled to various reservoirs at different temperatures. For this wide class of systems, we were able to obtain simple and useful closed-form expressions for both the Wigner entropy-production rate and the entropy flux, solely in terms of the second-order moments of the system. This is possible since we work with an overall harmonic Hamiltonian, which preserves Gaussianity. In addition, we could split the entropy-production rate and flux into contributions stemming from individual quasi-probability currents associated with each open decay channel, which enables one to identify the irreversibility generated by a single bath in the lattice. We also discussed how the internal coherent dynamics plays a central role in generating steady-state irreversibility, as it is the leading mechanism enabling energy transport across the network. All of the above should be understood as the leading (lowest-order) contribution in a perturbative expansion in the node-to-node interaction strengths (i.e., the limit of weak internal coupling), as it relies on the simple \textit{local structure} of the underlying quantum master equation. Finally, we used our framework to better understand the interplay between the various sources of irreversibility at play in diffusive heat conduction through a harmonic chain. We mimicked the anharmonicity required to establish the desired diffusive profile by adding auxiliary (self-consistent) reservoirs to our model. In turn, this allowed us to break down the total steady-state irreversible entropy production into a contribution due to the heat transport across the chain, and another one which can be interpreted as the entropic cost of maintaining a stationary diffusive transport.  

As we have shown, this approach offers several advantages, both practical and fundamental. 
From a practical standpoint, the fact that Gaussian states can be fully characterized by their covariance matrix allows one to readily access any entropic quantifier, even for very large system sizes. 
Secondly, from a fundamental point of view, our approach allows for a microscopic description of the problem of entropy production, enabling the identification of irreversible quasi-probability currents in phase space which are ultimately responsible for the emergence of irreversibility. Moreover, the description in terms of the Wigner function also allows us to take into account both thermal and quantum fluctuations, remaining valid even in the limit of very low temperatures, where the standard von Neumann approach becomes problematic. More generally, our framework can be readily adapted to quantify irreversibility in non-trivial energy-conversion processes (e.g., refrigeration) implemented on small-scale quantum heat devices. Indeed, networks of periodically driven and weakly interacting oscillators are a suitable platform to generate the required non-equilibrium states supporting useful continuous \textit{quantum-thermodynamic} cycles. 

\section*{Acknowledgments}

The authors are grateful to T. Tufarelli for useful comments. WTBM and GTL acknowledge the financial support from the S\~ao Paulo research foundation (FAPESP), under grants 2016/08721-7, 2017/50304-7, 2017/06323-7, and 2018/08211-4. LAC acknowledges funding from the European Research Council (StG GQCOP No. 637352). GTL acknowledges the financial support from the Brazilian funding agency CNPq (grant number INCT-IQ 246569/2014-0). MP acknowledges the H2020 Collaborative Project TEQ (grant agreement nr.  766900), the SFI-DfE Investigator Programme grant QuNaNet (grant 15/IA/2864), and the Leverhulme Trust project UltraQuTe. GTL and MP are funded by the FAPESP-QUB SPRINT program.

\appendix
\section{\label{app:cm}The covariance matrix formalism}

From $\bar{\op{R}} = (\op{a}_1, \op{a}_1^\dagger, \cdots, \op{a}_L,\op{a}_L^\dagger)^\mathsf{T}$, we may define the $ 2L\times 2L $ covariance matrix (CM) $ \mathsf{\mathsf{\Theta}} $ of our $ L $-node network as  
\begin{equation}
\Theta_{ij} = \frac{1} {2} \langle \{\op{R}_i,\op{R}_j^\dagger\}_+ \rangle - \langle \op{R}_i \rangle \langle \op{R}_j^\dagger \rangle,
\label{CM_theta}
\end{equation}
Hence, in our convention, the CM of system with $ L = 2 $ is 
\begin{equation}
\mathsf{\mathsf{\Theta}} = \begin{pmatrix}
\langle  \op{a}_1^\dagger  \op{a}_1 \rangle + \frac{1}{2} & \langle  \op{a}_1  \op{a}_1 \rangle & \langle  \op{a}_1  \op{a}_2^\dagger \rangle & \langle  \op{a}_1  \op{a}_2 \rangle \\[0.2cm]
\langle \op{a}_1^\dagger  \op{a}_1^\dagger \rangle  & \langle  \op{a}_1^\dagger \op{a}_1 \rangle + \frac{1}{2}  &  \langle  \op{a}_1^\dagger  \op{a}_2^\dagger \rangle & \langle \op{a}_1^\dagger \op{a}_2 \rangle \\[0.2cm]
\langle  \op{a}_1^\dagger  \op{a}_2 \rangle & \langle \op{a}_1  \op{a}_2  \rangle& \langle  \op{a}_2^\dagger  \op{a}_2 \rangle + \frac{1}{2} & 
\langle  \op{a}_2  \op{a}_2 \rangle \\[0.2cm]
\langle  \op{a}_1^\dagger  \op{a}_2^\dagger \rangle & \langle  \op{a}_1  \op{a}_2^\dagger \rangle & 
\langle  \op{a}_2^\dagger  \op{a}_2^\dagger \rangle & \langle  \op{a}_2^\dagger  \op{a}_2 \rangle + \frac{1}{2}
\end{pmatrix}.
\label{CM_N2}
\end{equation}
This ordering turns out to be more convenient for the problem at hand. Indeed, $ \mathsf{\Theta} $ can be readily decomposed in terms of the reduced covariance matrices 
\begin{subequations}
	\label{eq:defs_CS}
\begin{align}
	C_{ij} &\coloneqq \langle \op{a}_j^\dagger \op{a}_i \rangle - \langle \op{a}_j^\dagger \rangle \langle \op{a}_i\rangle, 
	\label{def_C}
	\\
	S_{ij} &\coloneqq \langle \op{a}_i \op{a}_j \rangle- \langle \op{a}_i \rangle \langle \op{a}_j \rangle. 
	\label{def_S}
\end{align}
\end{subequations}
One may then readily verify that 
\begin{equation}\label{app_fac}
\mathsf{\Theta} = \frac{\openone_{2L}}{2} + \mathsf{C} \otimes \s_+ \s_- + \mathsf{C}^\top \otimes \s_- \s_+ + \mathsf{S} \otimes \s_+ + \mathsf{S}^* \otimes \s_-,
\end{equation}
where $\openone_{2L}$ is the $2L$-dimensional identity matrix and $\sigma_{\pm}$ are raising and lowering spin-$1/2$ matrices. In the NESS of the problem discussed in Sec.~\ref{sec:chain}, $ S $ is identically zero, so that one only needs the reduced matrix $ \mathsf{C} $.

If the state of the system is Gaussian, the Wigner function \eqref{Wigner_definition} is completely determined by the CM and the vector of means $\bar{\mu} = \langle \bar{\op{R}} \rangle$, as
\begin{equation}\label{W_gaussian}
W(\bar{\alpha}) = \frac{1}{\pi^L \sqrt{\det{\mathsf{\Theta}}}} \exp{\left\lbrace - \frac{1}{2} (\bar{\alpha}- \bar{\mu})^\dagger \mathsf{\Theta}^{-1} (\bar{\alpha} - \bar{\mu})\right\rbrace}.
\end{equation}
In passing, it is also convenient to note that the inverse CM $\mathsf{\Theta}^{-1}$ can be written as 
\begin{equation}
\mathsf{\Theta}^{-1} = \frac{\openone_{2L}}{2} + '	
\mathsf{B} \otimes \s_+ \s_- + \mathsf{B}^\top \otimes \s_- \s_+ + \mathsf{P} \otimes \s_+ + \mathsf{P}^* \otimes \s_-,
\end{equation}
where $ \mathsf{B} = \big[\mathsf{C} - \mathsf{S}\,\big(\mathsf{C}^{-1}\big)^\top \mathsf{S}^* \big]^{-1} $ and $ \mathsf{P} = - \mathsf{C}^{-1}\,\mathsf{S}\, \mathsf{B}^\top $. In particular, if $\mathsf{S} = 0$, as in Sec.~\ref{sec:chain}, one simply gets $\mathsf{B} = \mathsf{C}^{-1}$.

\section{\label{app:lyap}Lyapunov equation}

Since the dynamics is Gaussian preserving, the state can be characterized at all times using only the CM $\mathsf{\Theta}$ and the mean vector $\bar{\mu}$, when starting from a Gaussian initial condition. In particular, the dynamics of the CM can be cast as the Lyapunov equation 
\begin{equation}\label{app_lyap}
\frac{\ud \mathsf{\Theta}}{\ud t} = \mathsf{W}\,\mathsf{\Theta} + \mathsf{\Theta}\,\mathsf{W}^\dagger + \mathsf{F},
\end{equation}
where $\mathsf{W}$ and $\mathsf{F}$ are $2L \times 2L$ matrices, which follow from the underlying master equation. In the case of Eq.~\eqref{master_eqn} 
\begin{equation}\label{app_W}
\mathsf{W} = - i \mathsf{H} \otimes \sigma_+ \sigma_ + + i \mathsf{H}^* \otimes \sigma_- \sigma_+ - \frac{\mathsf{\Gamma}}{2},
\end{equation}
where $\mathsf{\Gamma} = \text{diag}\,(\gamma_1, \gamma_1, \ldots, \gamma_L, \gamma_L)$ is a $2L\times 2L$ diagonal matrix with the specified diagonal elements and $\mathsf{H}$ is the matrix with elements $ H_{k,\ell} $ which determine the structure of the network. In turn, the matrix $ \mathsf{F} $ can be cast as $ \mathsf{F} \coloneqq \mathsf{f} + \mathsf{\Gamma}/2$, where 
\begin{equation}\label{app_F}
\mathsf{f} = \text{diag}\,(\gamma_1 n_1, \gamma_1 n_1, \ldots, \gamma_L n_L, \gamma_L n_L ).
\end{equation}

Since the master equation~(\ref{master_eqn}) does not spontaneously generate squeezing, it is possible to convert Eq.~\eqref{app_lyap} into two separate equations for the reduced CMs $C$ and $S$ [cf. Eqs.~\eqref{eq:defs_CS}]. To accomplish this, one simply needs to exploit the tensor structure of Eqs.~\eqref{app_W} and \eqref{app_F} together with that of Eq.~\eqref{app_fac}. This gives
\begin{subequations}
\begin{align}
\dfrac{\ud \mathsf{C}}{\ud t} &= i[\mathsf{C},\mathsf{H}] + \{\mathsf{C},\mathsf{\Gamma}\}_+ + \mathsf{f},
\label{dcdt}
\\
\dfrac{\ud \mathsf{S}}{\ud t} &= - i(\mathsf{S} \mathsf{H}^* + \mathsf{H} \mathsf{S}) + \{ \mathsf{S},\mathsf{\Gamma}\}_+.
\label{dsdt}	
\end{align}
\end{subequations}

Interestingly, we see that, while the equation for $\mathsf{C}$ acquires a clean structure, the equation for $\mathsf{S}$ becomes dependent on whether or not the matrix $\mathsf{H}$ is real or complex. In this respect, we note that the choice of phase in Eq.~\eqref{Hamiltonian1b} is particularly convenient, as it entails $\mathsf{H}^* = -\mathsf{H}$ and Eq.~\eqref{dsdt} acquires the same structure as Eq.~\eqref{dcdt}. We also remark that, while Eq.~\eqref{dcdt} has an inhomogeneous term $\mathsf{f}$, this is not present in Eq.~\eqref{dsdt}. As a result, in the NESS we get $\mathsf{S} = 0$ whereas $\mathsf{C}$ is the solution to
\begin{equation}\label{lyap_ness_C}
i[\mathsf{C},\mathsf{H}] + \{\mathsf{C}, \mathsf{\Gamma}\}_+ + \mathsf{f} = 0.
\end{equation} 

\section{\label{app:Pi_CM}Entropy-production rate}

The entropy production rate \eqref{eq:Pi_k} can be written in terms of the entries of the CM. We begin by substituting explicitly the current \eqref{currents_log} into Eq.~\eqref{eq:Pi_k}, which yields 
\begin{equation}
\Pi_k = \Phi_k - \gamma_k + \gamma_k\,\left(n_k+1/2\right) \int \ud \bar{\alpha }\,W\,\vert \pt_k \ln{W}\vert^2.
\label{eq:app:pi_n1}
\end{equation}
Next, we substitute the explicit formula \eqref{W_gaussian} for the Gaussian Wigner function in the logarithm and carry out the remaining integrals to obtain 
\begin{equation}\label{app_Pi_CM}
\Pi_k = \Phi_k - \gamma_k + \gamma_k\,\left(n_k+1/2\right)\,[\mathsf{\Theta}^{-1}]_{2k,2k}.
\end{equation}
Although this is not as simple as the expression for the entropy flux in Eq.~\eqref{Phi_k_2}, this closed formula for $ \Pi_k $ is equally useful.

\section{NESS from the model in Sec.~\ref{sec:chain}}
\label{app:CM_chain_NESS}

In this Appendix, we obtain the steady-state solution of Eq.~\eqref{M_model} from Sec.~\ref{ssec:model}. As discussed in Appendix~\ref{app:lyap}, this translates into solving an equation with the structure \eqref{lyap_ness_C}. In our specific case, the matrix $\mathsf{\Gamma}$ takes the form 
\begin{equation}
\mathsf{\Gamma} = \text{diag}\left(\gamma + \Gamma, \Gamma, \cdots, \Gamma, \gamma + \Gamma\right) \coloneqq \mathsf{\Gamma}_\text{r} + \Gamma \,\openone_{2L},
\end{equation}
where the subindex ``r'' refers to quantities of the physical reservoirs. Similarly, the elements of the matrix $\mathsf{f}$ in Eq.~\eqref{app_F} becomes
\begin{align}
\mathsf{f} &= \gamma\,\text{diag}\left(n_1,0, \cdots, 0,n_L\right) \nonumber\\
& + \Gamma\,\text{diag}\left(\langle \op{a}_1^\dagger \op{a}_1 \rangle, \langle \op{a}_2^\dagger \op{a}_2 \rangle, \cdots, \langle \op{a}_L^\dagger \op{a}_L \rangle\right) \nonumber\\
&\coloneqq \mathsf{f}_\text{r} + \Gamma \, \mathsf{C}_\text{d},
\end{align}
where $\mathsf{C}_\text{d}$ is a matrix containing only the diagonal elements of $\mathsf{C}$, with all other entries being zero. We thus see that the self-consistent reservoirs introduce the elements of $ \mathsf{C} $ directly into $\mathsf{f}$. Eq.~\eqref{lyap_ness_C} then becomes
\begin{equation}\label{app_lyap_asadian}
i[\mathsf{C},\mathsf{H}] + \{\mathsf{C}, \mathsf{\Gamma}_\text{r}\}_+ + \mathsf{f}_\text{r} + \Gamma\,\left(\mathsf{C}_\text{d} - \mathsf{C}\right) = 0.
\end{equation}
Interestingly, this is precisely the same equation that appears when one uses local dephasing to enforce diffusive heat conduction (cf., e.g., Ref.~\cite{Znidaric_2010,Asadian,Asadian2013_QHO}). We emphasize, however, that the NESS is \textit{different} in both models since dephasing is not Gaussianity-preserving, i.e., they only coincide up to the second-order moments.

The solution of Eq.~\eqref{app_lyap_asadian} was discussed in \cite{Znidaric_2010,Asadian,Asadian2013_QHO} and yields a tridiagonal matrix of the form
\begin{equation}
\mathsf{C} = \begin{pmatrix}
\langle \op{a}_1^\dagger \op{a}_1 \rangle & x &  &  & \\[0.2cm]
x  & \langle \op{a}_2^\dagger \op{a}_2 \rangle  &  x &  & \\[0.2cm]
& \ddots & \ddots & \ddots &  \\[0.2cm]
&   & x & \langle \op{a}_{L-1}^\dagger \op{a}_{L-1} \rangle & x \\[0.2cm]
&   &   & x & \langle \op{a}_L^\dagger \op{a}_L \rangle & \\[0.2cm]
\end{pmatrix}
\label{C_matrix}
\end{equation}
where $ x \coloneqq \langle \op{a}_k^\dagger \op{a}_{k+1}\rangle$ is given in Eq.~\eqref{corr_example} and the occupations are those of Eq.~\eqref{Ckk_example}.

\section{\label{app:separation_pis}Calculation of $\Pi_\text{r}$ and $\Pi_\text{sc}$ for the model in Sec.~\ref{sec:chain}}

Finally, we discuss how the above tools can be used to facilitate the computation of the individual entropy production rates for the real and the self-consistent reservoirs; namely Eqs.~\eqref{Pi_r_example} and \eqref{Pi_sc_example}. This is readily accomplished using Eq.~\eqref{app_Pi_CM}. For the contribution to the real reservoir we get 
\begin{equation}\label{app_Pi_r_example}
\Pi_\text{r} = \sum\nolimits_{k=1,L}\left[\Phi_k + \gamma\,\left(n_k+1/2\right)\,[\mathsf{\Theta}^{-1}]_{2k,2k}\right] - 2\gamma,
\end{equation}
where the entropy flux $\Phi_k$ is given by Eq.~(\ref{Phi_k_2}). Similarly, the entropy production rate due to the self-consistent reservoirs evaluates to
\begin{equation}\label{app_Pi_sc_example} 
\Pi_\text{sc} = \sum\nolimits_{k=1}^L\left[\Gamma\,\left(\langle \op{a}_k^\dagger \op{a}_k \rangle + 1/2\right)\,[\mathsf{\Theta}^{-1}]_{2k,2k}\right] - L\,\Gamma,
\end{equation}
were we have used the fact that the occupations of the self-consistent reservoirs are $\tilde{n}_k = \langle \op{a}_k^\dagger \op{a}_k \rangle$, so that the corresponding entropy fluxes $\tilde{\Phi}_k$ are identically zero. 
  
\bibliographystyle{apsrevfixedwithtitles} 
\bibliography{library}
\end{document}